\documentstyle{article}
\input {epsf}
\makeatletter
\def\Lm{\Lambda}
\newcommand{\beq}{\begin{equation}}
\newcommand{\eeq}{\end{equation}}
\newcommand{\beqy}{\begin{eqnarray}}
\newcommand{\eeqy}{\end{eqnarray}}

\def\cW{{\cal W}}

\catcode`\@=11
\def\numberbysection{\@addtoreset{equation}{section}
\def\theequation{\arabic{section}.\arabic{equation}}}
\def\appendix{\setcounter{section}{0}
        \def\thesection{Appendix \Alph{section}}
        \def\theequation{\Alph{section}.\arabic{equation}}}

\begin{document}
\begin{flushright}
hep-th/0011168
\end{flushright}
\renewcommand{\thefootnote}{\fnsymbol{footnote}}
\vspace{0.5in}
\begin{center}\Large{\bf 
Splitting of Heterogeneous Boundaries
in a System of\\
the Tricritical Ising Model Coupled to 2-Dim Gravity
}\\
\vspace{1cm}
\normalsize\ Masahiro ANAZAWA \footnote[2]
{E-mail address: anazawa@tohtech.ac.jp} 
\hspace{2mm}
and
\hspace{4mm} 
Atushi ISHIKAWA \footnote[3]{
E-mail address: ishikawa@kanazawa-gu.ac.jp}\\
\vspace{10mm}
$~^{\dagger}$
	{\it Tohoku Institute of Technology,
	Sendai 982-8577, Japan\\}
\vspace{5mm}
$~^{\ddagger}$ 
	{\it Kanazawa Gakuin University,
	 Kanazawa 920-1392, Japan \\}
\vspace{10mm}
\vspace{0.1in}
\end{center}
\renewcommand{\thefootnote}{\arabic{footnote}}
\setcounter{footnote}{0}
\vspace{0.4in}
\baselineskip 17pt
 
 We study disk amplitudes whose boundaries have heterogeneous matter states
 in a system of $(4,5)$ conformal matter coupled to 2-dim
 gravity.
 They are analysed by using the 3-matrix chain model 
 in the large $N$ limit.
 Each of the boundaries is composed of two
 or three parts with distinct matter states.
 From the obtained amplitudes, it turns out that each 
 heterogeneous boundary loop splits into
 several loops and
 we can observe properties in the splitting phenomena
 that are common to each of them.
 We also 
 discuss the relation to boundary operators.
\vspace{5mm}

\newpage

\baselineskip 17pt

It is well konwn that the
$(m,m+1)$ unitary comformal model coupled to 2-dim quantum gravity
can be described by matrix models. \cite{HI} -- \cite{DKK}
Microscopically, the $(m,m+1)$ model has $m-1$  matter degrees of freedom, 
which correspond to the points of the $A_{m-1}$ Dynkin diagram. \cite{Pasq}
The boundary of a 2-dim surface is one of the most
important objects
for considering a quantum theory of gravity.
In most cases, however, boundary conditions on matter configurations
are restricted to homogeneous ones
(except for in the cases discussed in Refs.~\cite{Staudacher} and \cite{SY}).
In Refs.~\cite{AIT} and \cite{AIT2} 
the present authors, together with a collaborator,
 examined a disk amplitude
whose boundary is heterogeneously composed of two arcs 
with different matter states
in the case of $(4, 5)$ conformal matter. 
We found that the original single loop with heterogeneous
matter states
changes its shape and that it splits into several loops
with homogeneous matter states. \cite{AIT2}

In this paper, 
similar disk amplitudes are examined
once again.
Here we also study the case in which a boundary
consists of three arcs, and 
find more complicated phenomena.
From these phenomena,
we can identify
common properties
of the loop splittings. 
A loop with heterogenous matter states is considerd to be related to 
one with homogeneous matter states 
on which some boundary operator \cite{ANA}, \cite{MMS}
is inserted.
We also discuss the relation 
between heterogeneous loop amplitudes
and boundary operators investigated
in Ref.~\cite{ANA}.
\\
 \\
%
\underline{Action and Critical Potentials}\quad
We study disk amplitudes for a system of $(4, 5)$ matter coupled to 
2-dim gavity
using the 3-matrix chain model. 
The action we start with is  
\begin{eqnarray}
S(A,B,C) & = & \frac{N}{\Lambda} 
                   {\rm tr} \left\{ U_1(A) + U_2(B) + U_1(C) - A B - B C
                      \right\}.
\label{3-matrix model Action}
\end{eqnarray}
Here $A$, $B$ and $C$ are $N \times N$ unitary matrix variables,
and $\Lm$ is the bare cosmological constant.
As critical potentials,
we choose
$
U_1(\phi)  =  \frac{111}{16} \phi - \frac{9}{4} \phi^2
                - \frac{1}{3} \phi^3
$ and
$
U_2(\phi)  =  - \frac{3}{4} \phi^2 - \frac{1}{12} \phi^3 
$ . 
These can be found using 
the orthogonal polynomial method. \cite{DKK},\cite{AIT2}
\\ \\
\underline{Schwinger-Dyson Equations}
\quad
Our aim is to examine the disk amplitudes
\beq
W_{AB}(p, q, \Lm)=
\sum_{n=0}^{\infty} \sum_{m=0}^{\infty} 
\frac{\Lm}{N} \left\langle {\rm tr}(A^n B^m) \right\rangle
 p^{-n-1} q^{-m-1}
\;,
\eeq
\beq
W_{AC}(p, r, \Lm)=
\sum_{n=0}^{\infty} \sum_{m=0}^{\infty} 
\frac{\Lm}{N} \left\langle {\rm tr}(A^n C^m) \right\rangle
 p^{-n-1} r^{-m-1}
\;,
\eeq
\beq
W_{ABC}(p, q, r, \Lm)=
\sum_{n=0}^{\infty} \sum_{m=0}^{\infty} \sum_{k=0}^{\infty}
\frac{\Lm}{N} \left\langle {\rm }tr(A^n B^m C^k) \right\rangle
 p^{-n-1} q^{-m-1} r^{-k-1}
\;
\eeq
and their continuum universal counterparts 
$w_{AB}(\zeta_A, \zeta_B, t)$, $w_{AC}(\zeta_A, \zeta_C, t)$
 and $w_{ABC}(\zeta_A, \zeta_B, \zeta_C, t)$
 in the large $N$ limit.
 Here $p$, $q$ and $r$ are bare boundary cosmological constants,
 $\zeta_A$, $\zeta_B$ and $\zeta_C$ are their renormalized
 counterparts, and $t$ is the renormalized cosmological constant.
Boundaries of these disks have
heterogeneous matter states.
Regarding
$w_{AB}(\zeta_A, \zeta_B, t)$ and $w_{AC}(\zeta_A, \zeta_C, t)$,
each of the boundary loops
 consists of two arcs with distinct matter states.
The boundary for
$w_{ABC}(\zeta_A, \zeta_B, \zeta_C, t)$ is
also composed of three arcs.
In Ref.~\cite{AIT2},  $W_{AB}(p,q,\Lm)$ and 
$W_{AC}(p,r,\Lm)$ are calculated, but 
the identification of the continuum universal part 
$w_{AC}(\zeta_A,\zeta_C,t)$ given there is
not correct. 
In this paper, we examine the amplitudes 
$W_{AB}(p,q,\Lm)$ and $W_{AC}(p,r,\Lm)$ once again, and we
calculate the more complex $W_{ABC}(p,q,r,\Lm)$.
Investigating them, we
discuss the loop configurations of the
heterogeneous boundaries.

The Shwinger-Dyson technique is useful
for our calculations.
In a manner similar to that used in Ref.~\cite{AIT2},
we obtain the following relevant Shwinger-Dyson equations:
\begin{eqnarray}
&&W_{A B}(p, q,\Lm) = \frac{(\frac{9}{2}+p)W_B(q,\Lm)
 + W^{(A)}_B(q,\Lm) + W_A(p,\Lm)}
                        { W_A(p,\Lm)-y(p)+q },
\nonumber\\
\qquad
&&W_{A C}(p, r,\Lm) = \frac{(\frac{9}{2}+p)W_C(r,\Lm)
 + W^{(A)}_C(r,\Lm) - W_{AC}^{(B)}(p,r,\Lm)}
                        { W_A(p,\Lm)-y(p) },
\nonumber\\
\qquad
&&W_{A C}^{(B)}(p, r,\Lm) = \frac{(\frac{9}{2}+p)W_C^{(B)}(r,\Lm)
 + W^{(AB)}_C(r,\Lm) - W_{AC}^{(B^2)}(p,r,\Lm)}
                        { W_A(p,\Lm)-y(p) },
\nonumber\\
\qquad
&&\frac{3}{2}W_{A C}^{(B)}(p, r,\Lm)
+\frac{1}{4}W_{AC}^{(B^2)}(p,r,\Lm)
+(p+r)W_{A C}(p, r,\Lm)
-W_A(p,\Lm)-W_C(r,\Lm)=0,
\nonumber\\
\qquad
&&W_{A B C}(p,q, r,\Lm) = 
\frac{W_{AC}(p,r,\Lm)+(\frac{9}{2}+p)W_{BC}(q,r,\Lm)
 + W^{(A)}_{BC}(q,r,\Lm)}
                        { W_A(p,\Lm)-y(p)+q },
\nonumber\\
\qquad
&&W_{B C}^{(A)}(q, r,\Lm) = 
\frac{(\frac{9}{2}+r)W_{B}^{(A)}(q,\Lm)
 + W^{(AC)}_{B}(q,\Lm)+W_C^{(A)}(r,\Lm)}
                        { W_C(r,\Lm)-y(r)+q }.
\label{SD}
\eeqy
Here $y(p)=\frac{111}{16}-\frac{9}{2}p-p^2$,
$
W_B^{(A^n C^k)}(q,\Lm)=
\sum_{m=0}^{\infty} 
\frac{\Lm}{N}\left\langle {\rm tr}(A^n B^m C^k)\right\rangle q^{-m-1}
$
and
$
W_{AC}^{(B^m)}(p,r,\Lm)=
\sum_{n=0}^{\infty} 
\sum_{k=0}^{\infty} 
\frac{\Lm}{N}\left\langle {\rm tr}(A^n B^m C^k)\right\rangle
p^{-n-1} r^{-k-1}
$,
etc.
Combining Eq.~(\ref{SD}) and other 
elementary Shwinger-Dyson equations in Ref.~\cite{AIT2} and 
using the $Z_2$ symmetry, it turns out that 
$W_{AB}(p, q,\Lm)$, $W_{AC}(p, r,\Lm)$
 and $W_{ABC}(p, q, r,\Lm)$ can be expressed
in terms of $W_{A}(p,\Lm)$, $W_{B}(q,\Lm)$ and 
$W_{C}(r,\Lm)$.
The explicit expressions, however, are somewhat complicated.
%
%
\\ \\
\underline{Results}
\quad
The continuum limit can be realized using the renormalization
$\Lm=35-\frac{5}{2}a^2 t$, $p=\frac{3}{2}a \zeta_A$, 
$q=2a \zeta_B$ and $r=\frac{3}{2}a \zeta_C$
with the lattice spacing $a$. \cite{GM}
The continuum universal parts of $W_{AB}(p,q,\Lm)$, $W_{AC}(p,r,\Lm)$ 
and $W_{ABC}(p,q,r,\Lm)$
can be obtained by using the following expressions: \cite{AIT2}
\beqy
W_A(p,\Lm)&=& y(p) + 2 \zeta_A a  \pm \frac{2}{3} w_A(\zeta_A, t) a^{5/4}
 + {\cal O}(a^{6/4}) \;,
\nonumber \\
W_C(r,\Lm)&=& y(r) + 2 \zeta_C a  \pm \frac{2}{3} w_C(\zeta_C, t) a^{5/4}
 + {\cal O}(a^{6/4}) \;,
\nonumber\\
W_B(q,\Lm)&=& z(q) + \frac{9}{2}  \pm  w_B(\zeta_B, t)
 a^{5/4}   + {\cal O}(a^{6/4}) \;.
\label{eq:scaling} 
\eeqy
Here $z(q) = - \frac{3}{2} q - \frac{1}{4} q^2$ and
\beqy
 &&w_A(\zeta, t)=w_B(\zeta, t)=w_C(\zeta, t) 
 =w(\zeta, t) \nonumber \\
 &&\qquad\qquad
=\left( \zeta +\sqrt{\zeta^2-t} \right)^{5/4}
+\left( \zeta -\sqrt{\zeta^2-t} \right)^{5/4}
\eeqy
are universal disk amplitides with homogeneous 
boundary matter states.
It must be pointed out that 
the ${\cal O}(a^{6/4})$ terms
in Eq.~(\ref{eq:scaling})
are not necessary for identifying
the leading univarsal parts of $W_{AB}(p,q,\Lm)$,
$W_{AC}(p,r,\Lm)$ and $W_{ABC}(p,q,r,\Lm)$.
The continuum universal amplitudes, therefore,
can be expressed in terms of $w_A(\zeta_A, t)$, 
$w_B(\zeta_B, t)$ and $w_C(\zeta_C, t)$.
After tedious calculations, we obtain
\beqy
w_{AB}(\zeta_A, \zeta_B, t) &=& 
\frac{1}{\zeta_A + \zeta_B}
	\left\{ w_A(\zeta_A,t)^2 
	+ \sqrt{2}w_A(\zeta_A,t) w_B(\zeta_B,t) 
	+ w_B(\zeta_B,t)^2 - 2 t^{5/4} \right\},
\label{AB3}
\\
\qquad
w_{AC}(\zeta_A, \zeta_C, t) &=& 
\frac{w_A(\zeta_A,t)-w_C(\zeta_C,t)}{\zeta_A-\zeta_C} 
	\left\{ 4t^{5/4} -w_A(\zeta_A,t)^2 - w_C(\zeta_C,t)^2 \right\},
\label{AC3} 
\\
w_{ABC}(\zeta_A,\zeta_B,\zeta_C,t) &=&
\frac{w_A(\zeta_A,t) - w_C(\zeta_C,t)}{\zeta_A - \zeta_C}
\left\{
\frac{2 t^{5/4}-w_A(\zeta_A,t)^2 }{\zeta_A + \zeta_B}
+
  \frac{2 t^{5/4}-w_C(\zeta_C,t)^2  }{\zeta_B + \zeta_C}
  \right\}
\nonumber \\
&-&
 \frac{1}{(\zeta_A + \zeta_B)(\zeta_B + \zeta_C)}
 \Bigl\{
 \sqrt{2} w_B(\zeta_B,t)^3
 + \sqrt{2} w_A(\zeta_A,t) w_B(\zeta_B,t) w_C(\zeta_C,t)
\nonumber \\
&+& \;
  w_A(\zeta_A,t) w_B(\zeta_B,t)^2 
  + w_C(\zeta_C,t) w_B(\zeta_B,t)^2 
  -2\sqrt{2} t^{5/4} w_B(\zeta_B, t)
  \Bigr\}
 \;.
\label{ABC3}
\eeqy
These expressions result
from the terms of order $a^{3/2}$, $a^{11/4}$
and $a^{7/4}$, respectively.\footnote
{In Ref.~\cite{AIT2} we obtained $w_{AC}$ from the term
of order $a^{5/2}$, which we believe is not universal.}
\\ \\
\underline{Loop Configulations} \quad
In order to study the loop configurations of the boundaries,
the inverse Laplace transformations of
(\ref{AB3})--(\ref{ABC3}) are useful.
We find the following inverse Laplace transformed forms:
\beqy
\cW_{AB}(\ell_A,\ell_B,t)&=&{\cal L_A}^{-1} {\cal L_B}^{-1} \left[ 
				w_{AB}(\zeta_A,\zeta_B,t)	\right]
\nonumber \\
\qquad
&=&\theta(\ell_A-\ell_B) (\cW_A*\cW_A)(\ell_A-\ell_B,t)
+ \theta(\ell_B-\ell_A) (\cW_B*\cW_B)(\ell_B-\ell_A,t)
\nonumber \\
\qquad
&&+ \sqrt{2}
	\int_0^{{\rm min}(\ell_A,\ell_B)}d\ell~ 
	\cW_A(\ell_A-\ell,t)\cW_B(\ell_B-\ell,t)
-2t^{5/4} \delta(\ell_A-\ell_B),
\label{cAB3}\\
\cW_{AC}(\ell_A,\ell_C,t)&=&{\cal L_A}^{-1} {\cal L_C}^{-1} \left[ 
				w_{AC}(\zeta_A,\zeta_C,t)	\right]
\nonumber \\
\qquad
&=&
\int_0^{\ell_A}d\ell~ 
	\cW_{\{AC\}}(\ell_C+\ell,t)(\cW_A*\cW_A)(\ell_A-\ell,t)
\nonumber \\
\qquad
&&+\int_0^{\ell_C}d\ell~ 
	\cW_{\{AC\}}(\ell_A+\ell,t)(\cW_C*\cW_C)(\ell_C-\ell,t)
\nonumber \\
&&
-4t^{5/4} \cW_{\left\{AC\right\}}(\ell_A+\ell_C),
\label{cAC3} 
\\
\cW_{A B C}(\ell_A, \ell_B, \ell_C)&=&
	{\cal L_A}^{-1} {\cal L_B}^{-1} {\cal L_C}^{-1} 
		\left[ w_{ABC}(\zeta_A,\zeta_B,\zeta_C,t) \right]
\nonumber \\
\qquad
&=&
 \theta(\ell_A - \ell_B) \int_0^{\ell_A - \ell_B}d\ell~ 
(\cW_A * \cW_A)(\ell_A - \ell_B - \ell,t) 
		\cW_{\left\{AC\right\}}(\ell_C + \ell,t) 
\nonumber \\
\qquad
&&+ \theta(\ell_C - \ell_B) \int_0^{\ell_C - \ell_B}d\ell~ 
(\cW_C * \cW_C)(\ell_C - \ell_B - \ell,t) 
		\cW_{\left\{AC\right\}}(\ell_A + \ell,t) 
\nonumber \\
\qquad
&&- \sqrt{2}~ \theta(\ell_B - \ell_A - \ell_C) 
	(\cW_B * \cW_B * \cW_B) (\ell_B - \ell_A - \ell_C,t)
\nonumber \\
\qquad
&&- \sqrt{2} \int_0^{\ell_A}d\ell~ \int_0^{\ell_C}d\ell'~
	\theta(\ell_B - \ell_A - \ell_C + \ell + \ell')
\nonumber \\
&&\qquad
	\times \cW_A(\ell,t) 
	\cW_B(\ell_B - \ell_A - \ell_C + \ell + \ell',t) \cW_C(\ell',t) 
\nonumber \\
\qquad
&&- \theta(\ell_B-\ell_C)\int_0^{{\rm min}(\ell_A,\ell_B-\ell_C)}d\ell~
	\cW_A(\ell_A-\ell,t) 
	(\cW_B*\cW_B)(\ell_B - \ell_C - \ell,t)
\nonumber \\
\qquad
&&- \theta(\ell_B-\ell_A)\int_0^{{\rm min}(\ell_C,\ell_B-\ell_A)}d\ell~
	\cW_C(\ell_C-\ell,t) 
	(\cW_B*\cW_B)(\ell_B - \ell_A - \ell,t)
\nonumber \\
&&-2t^{5/4} \Bigl[
	\left\{\theta(\ell_A-\ell_B)
	+\theta(\ell_C-\ell_B)\right\}
	\cW_{\left\{AC\right\}}(\ell_A+\ell_C-\ell_B)
\nonumber \\
	&&-\sqrt{2}\theta(\ell_B-\ell_A-\ell_C)\cW_B(\ell_B-\ell_A-\ell_C)
\Bigr].
\label{cABC3}
\eeqy
Here $\cW(\ell,t)$
represents the inverse Laplace transformed function of 
$w(\zeta,t)$,
that is,
$
\cW(\ell,t)
 ={\cal L}^{-1} \left[ w(\zeta,t) \right]
$.
The symbol $*$ represents convolution:
$(\cW*\cW)(\ell,t)=\int_0^{\ell}d~\ell' \cW(\ell',t)\cW(\ell-\ell',t)$.
We have also used the formula 
\[
{\cal L}_A^{-1}{\cal L}_C^{-1} 
\left[ \frac{w_A(\zeta_A,t)-w_C(\zeta_C,t)}{\zeta_A-\zeta_C}	
\right]
=- \cW(\ell_A+\ell_C,t)
= - \cW_{\{AC\}}(\ell_A+\ell_C,t)
\;. \]
In this expression,
the boundary of $\cW_{\{AC\}}$ consists of two arcs $\ell_A$
and $\ell_C$.

Now let us consider the geometrical configurations of 
Eqs.~(\ref{cAB3})--(\ref{cABC3}).
We refer to a part of the boundary which is composed of the matirx $A$
as ``boundary $A$'' or ``arc $A$" and so on.
The first term on the right-hand side of Eq.~(\ref{cAB3})
represents the configuration depicted in Fig.~1(a).
The entire region of the boundary $B$ bonds to the boundary $A$.
The second term represents the ($A \leftrightarrow B$) case.
Similarly, the third term corresponds to the case in Fig.~1(b).
Parts of boundaries $A$ and $B$ are stuck to each other.
In each case, the original loop splits into two loops with
homogeneous matter states, and they are linked by the bridge,
where parts of the arcs are completely stuck to
each other.
The fourth term represents the contribution from the special case
in which
the entire boundaries $A$ and $B$ are stuck completely.
The first term in Eq.~(\ref{cAC3}),
represents the configuration depicted in Fig.~2.
Two points on the boundary between the arc $A$ and $C$ bond to
the boundary $A$ simultaneously.
The second term corresponds to the case in which two such points stick to
the boundary $C$ simultaneously.
In these cases, the original loop splits into two homogeneous loops
and one heterogeneous loop, and they are connected at only one point.
The third term represents the contribution from the special case
in which the two homogeneous split loops shrink away.
Finally, the first term in Eq.~(\ref{cABC3})
represents the configuration depicted in Fig.~3(a).
The entire region of the boundary $B$ and 
the point on the boundary between the arc $A$ and $C$ are stuck to
the boundary $A$ simultaneously.
The second term represents the ($A \leftrightarrow C$) case.
In each case, the original loop splits into two homogeneous loops 
and one heterogeneous loop.
Similarly, the third, fourth and fifth terms correspond to
the cases in Figs.~3(b), (c) and (d), respectively.
The sixth term corresponds to the ($A \leftrightarrow C$) case of Fig.~3(d).
Two parts of boundary $B$ bond to the boundaries $A$ and $C$
simultaneously.
The point on the boundary 
between the arc $A$ and $C$ is not stuck to anything.
In these cases, the original loop splits into three homogeneous loops.
The seventh term represents the contribution from the special case
in which the two homogeneous split loops in the first or second term
shrink away.
The eighth term also comes from the special case
in which
the two homogeneous split loops in the third, fourth, fifth or sixth term
shrink away and one homogeneous split $B$ loop remains.
We must comment on the possibility that the terms 
corresponding to these special cases should be dropped 
from Eqs.~(\ref{cAB3})--(\ref{cABC3}).
This is due to the fact that
a shrinking loop has a finite lattice length
composed of the matrix $A$, $B$ or $C$, and the contirbution from 
such a part may turn out to be non-universal.

In these loop splitting processes,
the point on the boundary between
an arc $A$ and $C$ has some bond effect.
In fact, from Eqs.~(\ref{cAC3}) and (\ref{cABC3}), 
we can easily show
\beqy
\lim_{\ell_B \rightarrow 0}
\cW_{ABC}(\ell_A,\ell_B,\ell_C,t) 
=
\cW_{AC}(\ell_A,\ell_C,t).
\label{relation}
\eeqy
Such a point, therefore,
can be considered to be equivalent to an infinitesimal boundary $B$.
Microscopically, at that point, there must be a triangle
corresponding to the matrix $B$ which connects the $A$ and $C$ triangles.
The relation (\ref{relation}) is very natural
and we recognize such a point as an infinitesimal arc $B$.
From this consideration, we can obtain the following concise set of
properties, which are common in the loop splitting phenomenon:
\footnote{
The terms proportional to $t^{5/4}$ may be non-universal.
For simplicity, we found properties for the amplitudes
where such terms are dropped.
}
\\
1. Boundaries $A$ and $C$ cannot be bonded directly,
but a boundary $B$ can stick to either $A$ or $C$.\\
2. In this process, a point on the boundary
between an arc $A$ and $B$
or between an arc $A$ and $C$ must be on the boundary between bonded arcs
and separated arcs.\\
3. When two boundaries $B$
stick to a boundary $A$ or $C$ simultaneously, they bond to the same
kind of boundary.\\
4. In this case, 
a boundary $B$ does not form a  homogeneous split loop.\\
\\
\underline{Relation to Boundary Operators} \quad
We should discuss the relation to boundary opertors.
First let us consider the case of  ${\cal W}_{AB}(\ell_A, \ell_B,t)$.
When  $\ell_B$ goes to zero, the entire bondary
approches one on which the matter state is almost homogenous and
is different at only one point locally.
We can consider some boundary operator to be
inserted on a homogeneous loop with state $A$. 
We can consider ${\cal W}_{AC}(\ell_A, \ell_C,t)$ similarly
as $\ell_C$ approaches zero.
In these cases, we can easily obtain 
\beqy
\lim_{\ell_B \to 0} {\cal W}_{AB}(\ell_A, \ell_B,t)
&=&({\cal W}_A * {\cal W}_A )(\ell_A,t)
\; ,
\label{eq:limit1}\\
\qquad
\lim_{\ell_C \to 0} {\cal W}_{AC}(\ell_A, \ell_C,t)
&=&({\cal W}_A * {\cal W}_A * {\cal W}_A)(\ell_A,t)
\; .
\label{eq:limit2}
\eeqy
From Eq.~(\ref{eq:limit1}),
we see that the insetrtion of the corresponding boundary operator
has the effect of splitting the original loop  
into two loops.
Similarly, the  boundary operator
corresponding to Eq.~(\ref{eq:limit2})
has the effect of splitting a loop  
into three loops.
In Ref.~\cite{ANA}, similar phenomena are discussed.
A system of $(m, m+1)$ conformal matter coupled to 2-dim gravity
has an infinite number of scaling operators.
They are classified into two groups.
In one of them, the scaling opertors are gravitationally dressed primary
operators of the $(m, m+1)$ model and their gravitaional decendants.
In the other group, the scaling operators are considered to be
boundary operators, which have the effect of splitting a loop 
into several loops. \cite{ANA}
We believe that it is natural to identify the 
boundary operators corresponding to Eqs.~(\ref{eq:limit1}) 
and (\ref{eq:limit2}) with 
the boundary operators $\widehat{B}_2=\widehat{\sigma}_{2(m+1)}$ 
and $\widehat{B}_3=\widehat{\sigma}_{3(m+1)}$ 
discussed in Ref.~\cite{ANA}.
\\

When matter states are heterogeneous on a boundary,
the shape of an original single loop changes,
and it splits into several loops.
This phenomenon was first pointed out in Ref.~\cite{AIT2},
in which each of the original loops consists of two parts
that have different matter states.
This phenomenon
could not be seen if we only considered a homogeneous boundary.
In this paper, we found more complex phenomena
for the case in which the boundary is composed of three parts.
From these amplitudes,
we found several common properties
in the loop splitting phenomenon.
We also pointed out a relation to boundary operators.
The properties 3 and 4 discussed above, however,
are, at this time, merely phenomenological.
We speculate that the mechanisms which underly the obtained properties
will be made clear by investigating the splitting phenomenon more deeply.

\begin{center}\section*{Acknowledgements}
\end{center}

We would like to express our gratitude to Professor M. Ninomiya
for warmhearted encouragement.
We are grateful to Professor H. Kunitomo for careful
reading of the manuscript.
Thanks are also due to members of YITP, where one of authors (A.I.)
stayed several times
during the completion of this work.
\\


\begin{figure}
\begin{center}
  \epsfbox{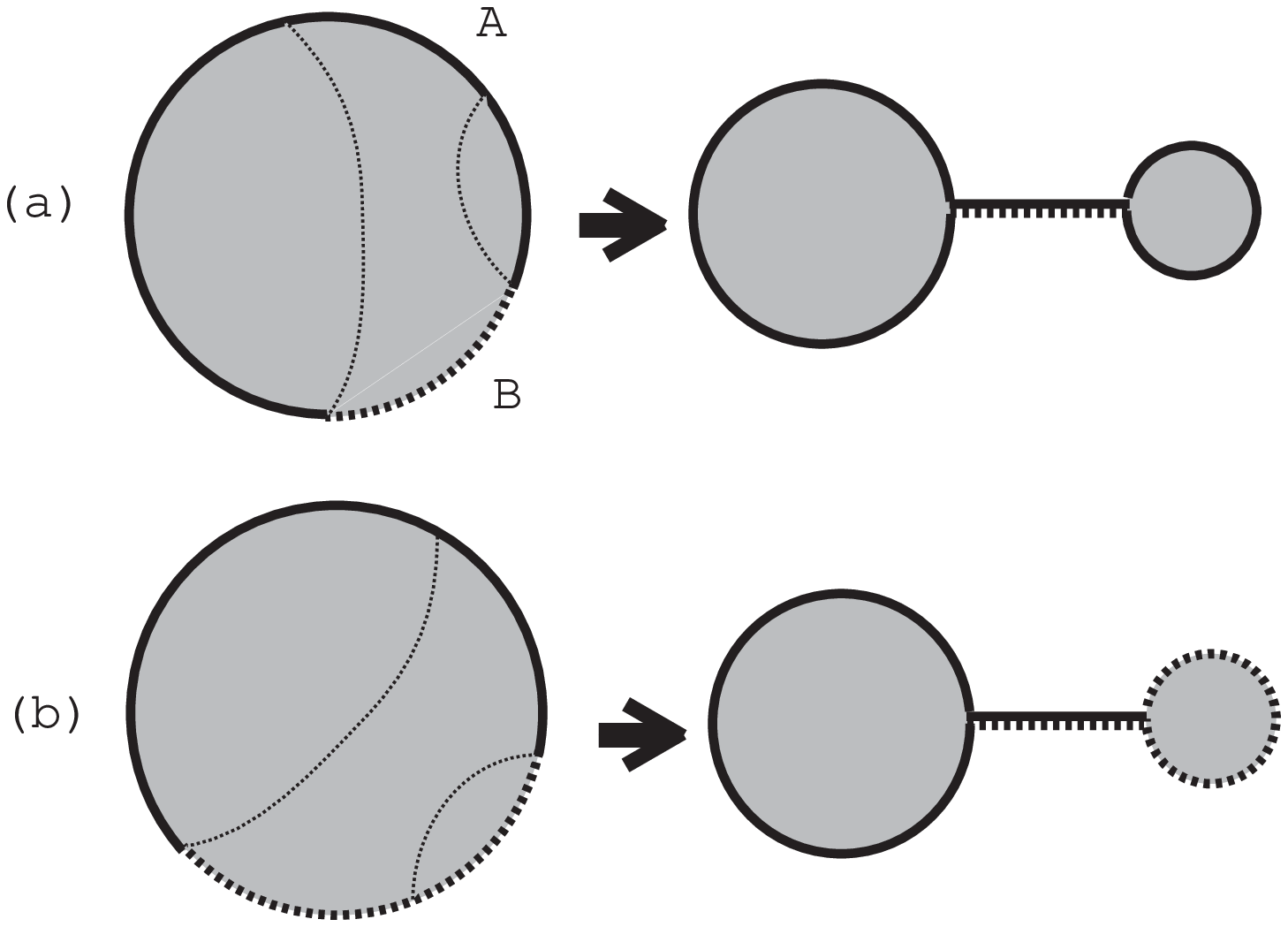}
  \caption{Due to the sticking of two different kinds of boundaries,
           the original loop  splits into two loops with
           homogeneous matter configurations.}
  \label{fig1}
\end{center}
\begin{center}
  \epsfbox{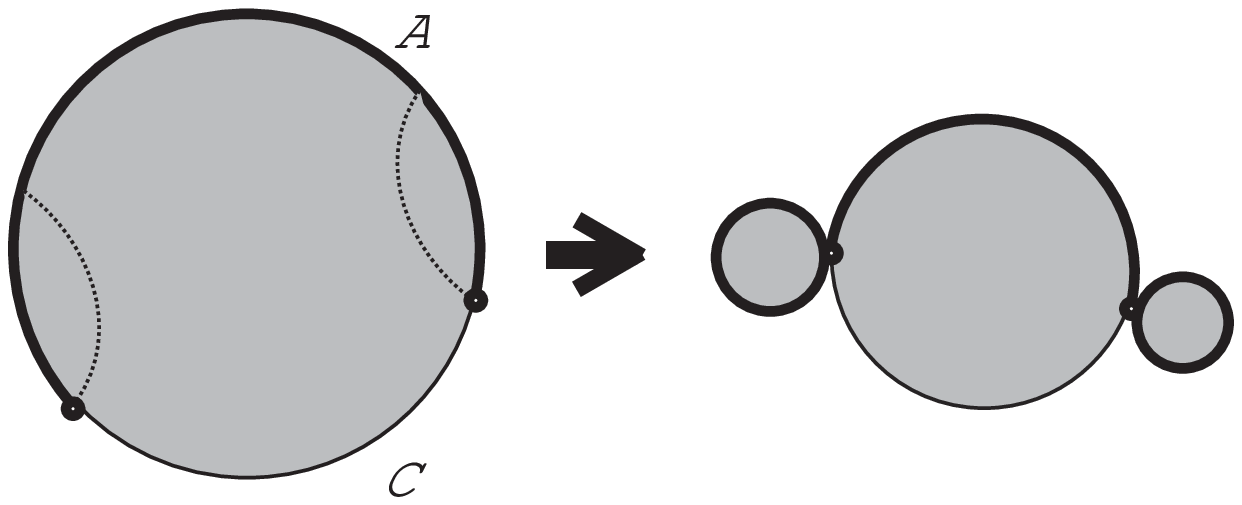}
  \caption{The original loop, composed of two different parts of a boundary,
	   splits into two homogeneous loops and one heterogeneous loop.}
  \label{fig2}
\end{center}
\end{figure}
\begin{figure}
\begin{center}
  \epsfysize=20cm
  \epsfbox{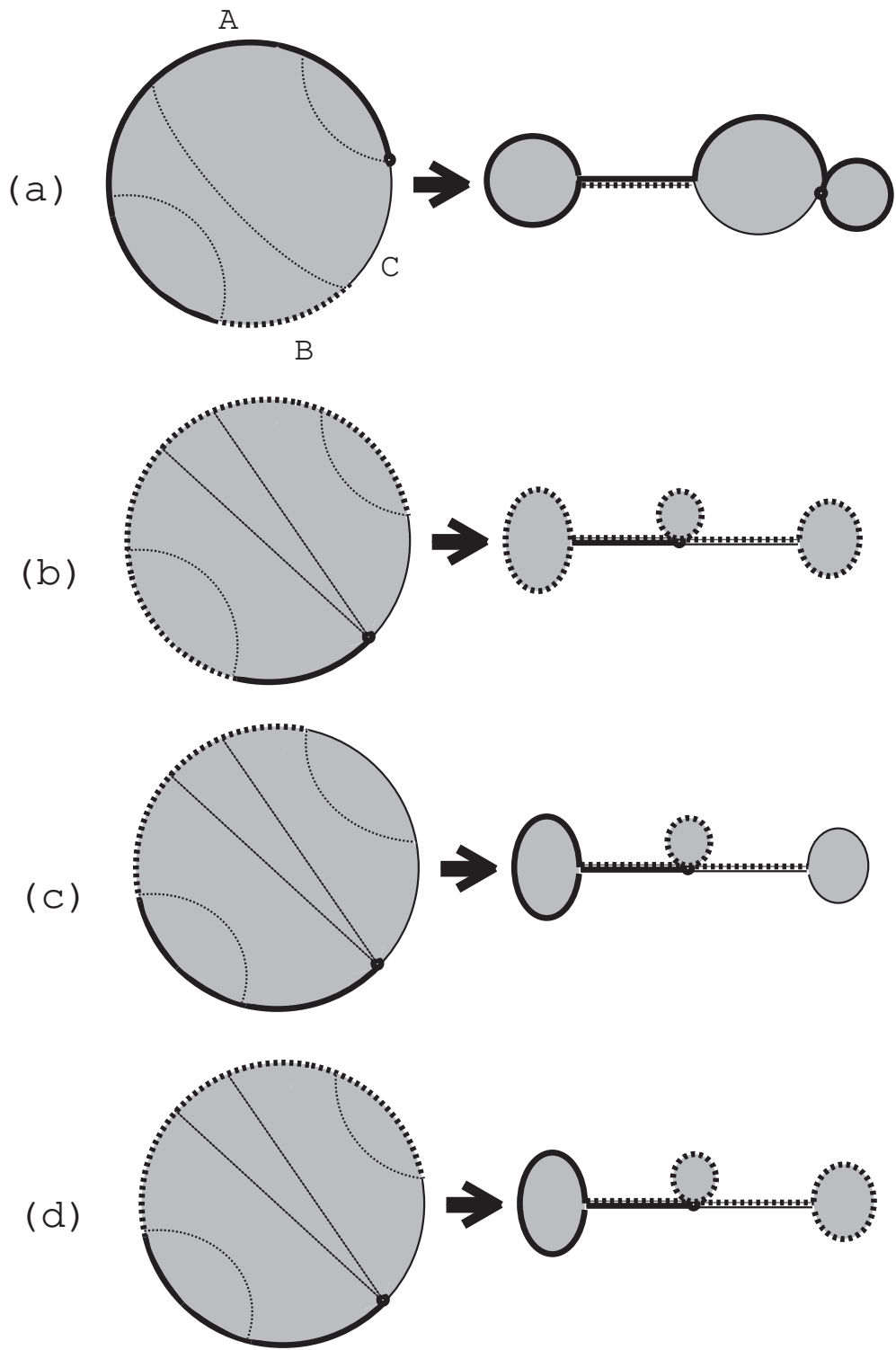}
  \caption{The original loop, composed of three different parts of a boundary,
	   splits into three loops.}
  \label{fig3}
\end{center}
\end{figure}

\end{document}